\documentclass{aastex63}
\graphicspath{{./}{figures/}}
\received{March, 2025}
\revised{December, 2025}
\accepted{December, 2025}
\submitjournal{PSJ}

\shorttitle{}
\begin{document}

\title{Retrograde predominance of small saturnian moons reiterates a recent retrograde collisional disruption}

\author[0000-0002-4637-8426]{Edward Ashton}
\affiliation{Institute of Astronomy and Astrophysics, Academia Sinica, No.1, Sec. 4, Roosevelt Road, Taipei 10617, Taiwan}

\author[0000-0002-0283-2260]{Brett Gladman}
\affiliation{Dept.~of Physics and Astronomy, University of British Columbia, Vancouver BC, Canada}

\author[0000-0003-4143-8589]{Mike Alexandersen}
\affiliation{Center for Astrophysics $|$ Harvard \& Smithsonian, 60 Garden Street, Cambridge, MA 02138, USA}

\author[0000-0003-0407-2266]{Jean-Marc Petit}
\affiliation{Université Marie et Louis Pasteur, CNRS, Institut UTINAM (UMR 6213), OSU
THETA, F-25000 Besançon, France}

%% Note that the \and command from previous versions of AASTeX is now
%% depreciated in this version as it is no longer necessary. AASTeX 
%% automatically takes care of all commas and "and"s between authors names.

%% AASTeX 6.2 has the new \collaboration and \nocollaboration commands to
%% provide the collaboration status of a group of authors. These commands 
%% can be used either before or after the list of corresponding authors. The
%% argument for \collaboration is the collaboration identifier. Authors are
%% encouraged to surround collaboration identifiers with ()s. The 
%% \nocollaboration command takes no argument and exists to indicate that
%% the nearby authors are not part of surrounding collaborations.

%% Mark off the abstract in the ``abstract'' environment. 
\begin{abstract}

We report the discovery and careful orbital determination of 64 new irregular moons of Saturn found in images taken using the Canada-France-Hawaii Telescope from 2019-2021, bringing the total number of saturnian irregulars to 122. By more than doubling the sample of saturnian irregular moon orbits, including pushing to smaller sizes, we can now see finer detail in their orbital distribution. We note the emergence of potential subgroups associated with each of Siarnaq and Kiviuq within the Inuit group.

We find that in the inclination range 157--172 degrees the ratio of smaller moons (diameters less than 4 km) to larger moons (diameters greater than 4 km) is significantly larger than that of any other inclination range in the retrogrades. We denote this subset of the Norse group as the Mundilfari subgroup after its largest member. The incredibly steep slope of the Mundilfari subgroup's size distribution, with a differential power law index of q = 6, strengthens the hypothesis in \citet{Ashton2021} that this subgroup was created by a recent catastrophic collision, $<10^8$~yr ago.

%Of the new discoveries, a vast majority (54/64) have retrograde orbits.
  
\end{abstract}

%% Keywords should appear after the \end{abstract} command. 
%% See the online documentation for the full list of available subject
%% keywords and the rules for their use.

%% From the front matter, we move on to the body of the paper.
%% Sections are demarcated by \section and \subsection, respectively.
%% Observe the use of the LaTeX \label
%% command after the \subsection to give a symbolic KEY to the
%% subsection for cross-referencing in a \ref command.
%% You can use LaTeX's \ref and \label commands to keep track of
%% cross-references to sections, equations, tables, and figures.
%% That way, if you change the order of any elements, LaTeX will
%% automatically renumber them.
%%
%% We recommend that authors also use the natbib \citep
%% and \citet commands to identify citations.  The citations are
%% tied to the reference list via symbolic KEYs. The KEY corresponds
%% to the KEY in the \bibitem in the reference list below. 

\section{Introduction} \label{sec:intro}

All four giant planets in the Solar System have moons that are thought to be planetesimals that were captured by their host planet during the final stages of planet formation.
These moons, known as irregular moons, can be distinguished from their regular counterparts (moons that formed around their host planet as the planet was forming) by their orbits. 
The semi-major axes of irregular moons are typically much larger than those of regular moons, hundreds of planetary radii versus a few to tens of planetary radii.
Additionally, irregular moons have moderate to large orbital eccentricities and a variety of inclinations, whereas the regular moons have near-circular orbits that are situated around the equator of the host planet. See \citet{Jewitt&Sheppard2005}, \citet{Nicholson2008}, and \citet{Denk2018} for review articles covering saturnian irregular moons.

Phoebe was the first irregular moon of Saturn discovered back in 1898.
Since then, excluding the moons found in this work, just two surveys have been able to discover irregular moons around Saturn. 
A bit more than 100 years after Phoebe's discovery, \citet{Gladman2001} found 12 irregular moons from images taken in 2000 and 2001 using multiple telescopes with wide-field capabilities. 
Their orbits yielded the first evidence of inclination groupings for saturnian irregulars.

The other successful survey at finding saturnian irregulars used the Subaru Telescope between 2004 and 2007.
Overall, this search was able to discover and track 45 new moons. 
This survey was eventually outlined in \citet{Sheppard2023}. %[Added instead of:]
%Although there is currently no publication that describes this survey,
The reviews of \citet{Nicholson2008} and \citet{Denk2018} contain discussions of 25 of the 45 moons from this survey. Combining Phoebe with the moons found in the two aforementioned surveys, a total of 58 saturnian irregulars were known by 2019.

%The orbits of irregular moons can be significantly tilted relative to the orbital plane of the host planet, which causes large oscillations in both inclination and eccentricity due to solar perturbations.
Strong solar perturbations on these distant moons cause periodic oscillations in both inclination and eccentricity, taking $\sim$500~yr for saturnians.
Thus, time-averaged orbital elements are a more stable measure of an irregular moon's orbit than the osculating orbital elements;
these time-averaged orbital elements of irregular moons can be calculated using N-body simulations to monitor the evolution of each of the moon's orbital elements over time (see \citet{Nesvorny2003,Turrini2008,Jacobson2022}). Simulations need to be integrated over many oscillations to obtain accurate time-averaged elements. Unless otherwise stated, anytime semi-major axis ($a$), eccentricity ($e$), or inclination ($i$) of an irregular moon is mentioned hereafter we refer to the time-averaged value.

Two of the saturnian irregular groups are direct in nature ($i < 90^\circ$, also known as prograde): the Inuit group, with inclinations relative to the ecliptic of $\approx 48^\circ$, and the Gallic group, with inclinations of $\approx 38^\circ$. 
Both of the direct irregular groups were not well populated, with only six members each pre-2021, and have inclination spreads of only a few degrees.
The retrograde ($i > 90^\circ$) Norse group, in contrast, is much more dispersed, with inclinations ranging from 136$^\circ$ to 178$^\circ$, and contained a vast majority of the saturnian irregulars with 46 members. 

The $10^\circ$ separation between the Gallic and Inuit groups is a factor of two smaller than the up to 20$^\circ$ variation in inclination that a saturnian irregular can experience \citep{Nesvorny2003,Turrini2008,Jacobson2022}, which shows why calculating time-averaged orbital elements is vital for studying the orbital distribution of irregular moons. 
A good example is Saturn LX, with a current inclination of 44.4$^\circ$ it was initially thought that the moon was a member of the Inuit group. However, when using the time-averaged inclination 38.6$^\circ$, it was evident that Saturn LX belongs to the Gallic group \citep{Jacobson2022}.

It was initially thought that a single irregular moon group was created by the break up of a larger moon, like the collisional families seen in the asteroid belt.
However, some groups are too dispersed to be explained by a single collision \citep{Grav2003,Nesvorny2003,Turrini2008}.
A likely remedy is either multiple independent collisions, second-generation collisions between fragments of the initial collision, or a mixture of the two.
Over the years there have been many attempts to divide the groups 
into subgroups/pairs of saturnian irregulars \citep{Grav2003,Grav&Bauer2007,Turrini2008,Denk2018,Ashton2021}. However, these studies had only a small fraction (less than a third) of the currently known saturnian irregulars at their disposal.

Although there is currently no consensus on the exact mechanism by which the initial population of irregular moons was captured, multiple theories have been proposed. 
The most popular are gas drag by an extended envelop and three-body interactions (either the planet capturing one member of a passing binary, or planetesimal capture during a planet-planet encounter). 
Refer to \citet{Nicholson2008} for a detailed discussion of the proposed capture scenarios.
Knowing the orbital distribution of the initial irregular moon population would provide constraints on the capture mechanism. However, this is complicated by the fact that the irregular moon populations have likely undergone significant collisional alteration over the age of the Solar System \citep{Bottke2010}. 
Therefore, understanding the collisional history of an irregular moon system allows one to better constrain the initial population and thus the capture process, realising that only traces of the original captures are currently still present. %Is this the right place for this paragraph?

This paper details the results of a search for irregular moons of Saturn using the Canada-France-Hawaii Telescope (CFHT) from 2019 to 2021, and an analysis of the saturnian irregular moon population.
A summary of the observations and the search method can be found in \autoref{sec:Obs}. 
The 64 new moons discovered as a result of the survey are described in \autoref{sec:Moons}.
Analysis of the orbital distribution is found in \autoref{sec:Groups}. A look at the size distribution of groups/subgroups is presented in \autoref{sec:sizedist}. The case for the Mundilfari subgroup being a recent collisional family is detailed in \autoref{sec:col}.
Finally, we give concluding remarks in \autoref{sec:con}. 

\section{Observations}
\label{sec:Obs}

In the first 4 nights of July, 2019 CFHT acquired images of one of two fields near Saturn (one east and one west of the planet, with offsets shown in Fig.~\ref{fig:tracking}). Each night contained 44 sequential wide-band (gri) images that were acquired over an approximately 3 hour period. The 44 205-sec images from a night were combined using the shift and stack technique \citep{Gladman1998,Kavelaars2004,Ashton2021} to detect moving irregular moons of Saturn down to diameters, $D$, of approximately 3~km ($m_w=26.3$). A total of 120 objects moving at Saturn-like rates were detected, of which 45 were eventually linked to previously designated moons. 
This search and a debiased size distribution over the entire moon system is detailed in \citet{Ashton2021}.
At this time, even the orbital sense (i.e. retrograde versus direct) of most of the orbits were not known.

In order to track the remaining 75 candidate moons, to confirm their moon status, and to obtain their orbits, additional observations were acquired.
The original plan was to observe both fields in the 3 dark runs following the initial July 2019 observations. However, due to various reasons, images were obtained only in the end of August/ start of September dark run. Additionally, these nights only went to a depth of $m_w$ $\sim$ 26 due to a combination of higher stellar density, extinction due to clouds, and poorer image quality. 

Due to inadequate followup observations taken in 2019, more images were acquired in 2020. Both fields were re-imaged on two nights each in June, and then one night each a month later in July. 
The west field got a bonus, but partial, sequence of 29 images in August.
A final sequence (for each field) was acquired in July 2021. 
The same filter, the same field offsets from Saturn, and the same shift and stack technique used for the July 2019 observations were also used in all of the follow-up observations.

%\begin{deluxetable*}{lccccccc}
%  \tabletypesize{\scriptsize}
%  \setlength{\tabcolsep}{0.1in}
%  \tablecaption{A summary of the observations. \label{tab:sum}}
%  \tablewidth{0pt}
%  \tablehead{ \colhead{Field} & \colhead{Number of} & \colhead{Date} & \colhead{Average Seeing} & \colhead{Average Airmass}  & \colhead{Comment}  \\
%  \colhead{} & \colhead{Images} & \colhead{(UTC)}  & \colhead{}  & \colhead{} & \colhead{}}
%\startdata
%East &  44 &  2019-07-01  &  &  & - \\
%East &  44 &  2019-07-02  &  &  & - \\
%West &  44 &  2019-07-03  &  &  & - \\
%West &  44 &  2019-07-04  &  &  & - \\
%West &  46 &  2019-08-22  &  &  & - \\
%East &  45 &  2019-08-23  &  &  & - \\
%East &  52 &  2019-08-26  &  &  & Extinction due to clouds \\
%East &  44 &  2019-08-27  &  &  & Extinction due to clouds \\
%West &  47 &  2019-08-28  &  &  & - \\
%East &  44 &  2019-09-03  &  &  & Extinction due to clouds \\
%East &  35 &  2020-06-16  &  &  & Not a complete sequence \\
%East &  44 &  2020-06-24  &  &  & - \\
%West &  44 &  2020-06-27  &  &  & - \\
%West &  44 &  2020-06-28  &  &  & - \\
%East &  44 &  2020-06-29  &  &  & - \\
%West &  46 &  2020-07-20  &  &  & - \\
%East &  44 &  2020-07-24  &  &  & - \\
%West &  29 &  2020-08-16  &  &  & Not a complete sequence \\
%East &  44 &  2021-07-08  &  &  & - \\
%West &  44 &  2021-07-09  &  &  & - \\
%\enddata
%\end{deluxetable*}

\section{Detections}
\label{sec:Moons}

\begin{figure}[ht]
\plotone{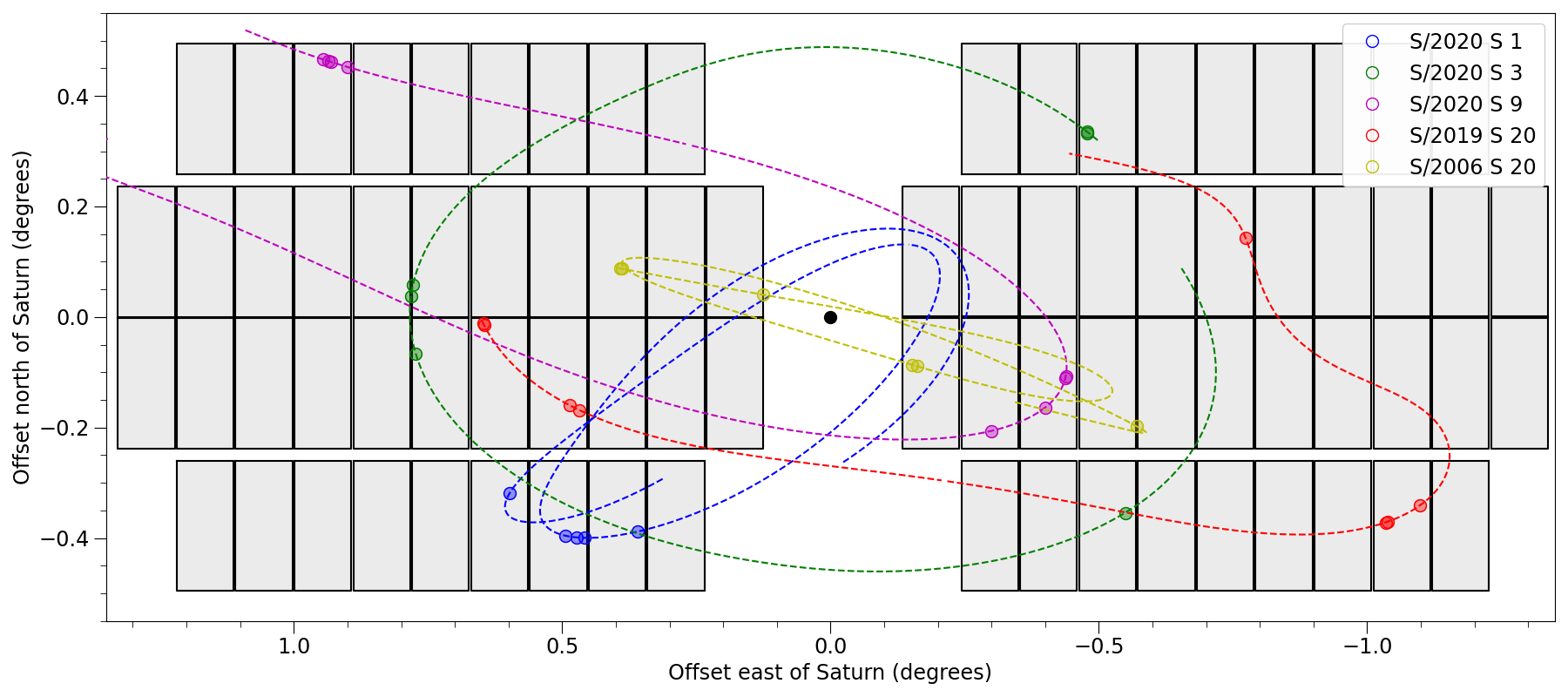}
\caption{On-sky plot of the two fields used for this survey relative to Saturn (grey rectangles). All measurements of 5 of the 64 new moons detected are shown (circles) along with the best fit orbit (dashed lines).
This sub-sample illustrates the challenges
of multi-year linking.  (Note that the apparent non-closure of
the orbits is due to the projected saturnocentric positions being observed from the moving Earth.)
}
\label{fig:tracking}
\end{figure}

Over all images in our survey we were able to observe all of the 58 previously known saturnian irregulars except for one, Bestla.
Bestla's large orbital inclination and node location resulted in it spending a large fraction
of its time during these observations nearly due north or south of Saturn, which our fields did not cover
(Fig.~\ref{fig:tracking});
because other moons of comparable inclinations were both recovered and newly detected, this must partly be a timing issue of where Bestla was at the time.
All of the previously known moons now have 14-year observational arcs at the minimum\footnote{Even though we did not observe Bestla, another team measured it in 2015, providing a 14-year observational arc.}, and thus possess very high-precision orbits.
Our survey produced 64 new irregular moons of Saturn, details of which are provided below.

\subsection{Finding new moons}

The sparse observations made linking the candidate moons challenging.
The method that produced a vast majority of the new discoveries involved linking candidate moons in the 2020 opposition through linear extrapolation to obtain a preliminary orbit from a month arc and then extrapolating this orbit in an attempt to find linkages in the 2019 and 2021 observations. 
Although linking candidate moons within the 2020 opposition was straightforward, the number of observations in 2020 is so little that most of the orbits were relatively unconstrained. As such, the orbit was able to be significantly altered
while still providing a good fit, resulting in many candidate linkages that had to be checked.

The rest of the new moons were confirmed using the same method stated in the previous paragraph but starting with the 2019 observations.
However, this method had a couple of setbacks. One being the vast departure of the actual position of the moons from the linear extrapolation over two months, making it challenging to find the July 2019 candidates in the Aug/Sept 2019 observations.
The other issue is the relatively shallow depth of the Aug/Sept 2019 images, as detailed in \autoref{sec:Obs}.
This meant that only the brightest moon candidates in the July 2019 images were able to be found in the Aug/Sept 2019 data, and thus tracked to the following two oppositions.

In total, 64 new irregular moons of Saturn have been discovered using our data set. 
One of these, S/2019 S 1 whose very small semi-major axis results in being visible outside Saturn's glare for only a small fraction of the time, was announced in 2021 \citep{Ashton2022}; the rest were announced in May 2023.
Additionally, we were able to detect 50+ more moon candidates that we were unable to link together due to an insufficient number of detections.

%------------------------------------ FIGURE 1 -------------------------------
%------------------------------ end Figure 1 --------------------------------------

\subsection{Linking back to past surveys}

Like this survey, the 2003-2007 survey using Subaru \citep{Sheppard2023} detected many moon candidates that were unable to be connected into a long enough arc to produce confirmed moons. 
The precision of our multi-year orbits of the new moons is good enough that the Minor Planet Center was able to link back the majority of our linked moons to these previous candidates. 
Of the 64 new moons, 42 were matched to candidates found by Subaru from 2004--2007, ranging from a single night of observations to seven nights spread over multiple oppositions. 
If a solar system object is detected on at least two different nights in a single opposition, then the year of the first night becomes the discovery year. As such, 33 of the new moons ultimately have 2004--2007 as their discovery year and have one of those years as part of their temporary designation (see \autoref{tab:orbit}). 

\subsection{July 2019 detections}

Of the 75 moon candidates found in July 2019 that were not linked to a previously known moon,
44 are linked to the new discoveries,
leaving 31 candidates from the July 2019 data that are still not yet linked to a confirmed moon. 
Most of these remaining candidates are faint, with only nine being above the characterisation limit (50\% detection efficiency) of the CCD they were found on.
An additional 12 more new moons were detected in the July 2019 images. These moons were missed in the original search but were found from high-precision orbits obtained by linking 2020 and 2021 observations.

\section{Groups}
\label{sec:Groups}

For our analysis of the orbital distribution of the saturnian irregular moons we used the time-averaged orbital elements calculated by JPL\footnote{https://ssd.jpl.nasa.gov/sats/elem/$\#$refs}.
Their methods are detailed in \citet{Jacobson2022}, where they simulated the orbits of the 58 irregular saturnian moons known prior to this work, plus S/2019 S 1, over a 5,000~year period. All inclinations are relative to the ecliptic except for Phoebe, which is relative to the Laplace plane.

\subsection{Size}
\label{sec:size}

Before analysing the saturnian irregular moon population as a whole,
we used Equation 5 from \citet{Petit2008},

\begin{equation}
    H = m_{\sun} - 2.5 \log\left(\frac{p r^2}{2.25\times10^{16}}\right) \; ,
    \label{eq:radtoH}
\end{equation}

\noindent
to estimate each moon's size,
where $m_{\sun}$ is the Sun's apparent magnitude, and $H$, $p$, and $r$ are the Solar System absolute magnitude, geometric albedo, and radius in km of the moon, respectively. 
For this equation to be valid, the values for $H$, $p$, and $m_{\sun}$ all need to be in the same filter.
We will use the V-band filter for our size calculations since that is the filter the Minor Planet Center uses for their values for $H$ of irregular moons\footnote{https://www.minorplanetcenter.net/iau/NatSats/NaturalSatellites.html}.
For $m_{V,\sun}$ we use a value of -26.77, which comes from $M_{V,\sun}=4.80$ \citep{Willmer2018}.
Since only a few of the saturnian irregulars have measured albedos we have decided to go with a single value of $p_V = 0.06$, which is the median value of the three saturnian irregular moons with measured albedos, $0.081 \pm 0.02$ for Phoebe \citep{simonelli1999}, $0.06 \pm 0.03$ for Albiorix, and $0.050 \pm 0.017$ for Siarnaq \citep{Grav2015}.
Substituting in the values for $m_{V,\sun}$ and $p_V = 0.06$, and turning $r$ into $D$ (diameter), \autoref{eq:radtoH} becomes
\begin{equation}
    \left(\frac{D}{2~km}\right)^2 =  10^{0.4(17.2-H_V)}
    \label{eq:abstoradV}
\end{equation}

\noindent
It should be noted that using a single albedo value means that the slope of the size distribution is independent of the albedo value. Thus, our choice of albedo has little bearing on our analysis.

\begin{deluxetable*}{lcccccccc}
  \tabletypesize{\scriptsize}
  \setlength{\tabcolsep}{0.1in}
  \tablecaption{A list of the irregular moons of Saturn with their semi-major axis ($a$), eccentricity ($e$), inclination ($i$), orbital period ($P$), V-band absolute magnitude ($H_V$), the group they belong to, and the subgroup we have assigned to them (if any). Orbital elements are taken from JPL\footnote{https://ssd.jpl.nasa.gov/sats/elem/$\#$refs} and $H_V$ from MPC\footnote{https://www.minorplanetcenter.net/iau/NatSats/NaturalSatellites.html}. All inclinations are relative to the ecliptic except for Phoebe, which is relative to the Laplace plane. \label{tab:orbit}}
  \tablewidth{0pt}
  \tablehead{ \colhead{Name/} & \multicolumn{2}{c}{$a$} & \colhead{$e$} & \colhead{$i$}  & \colhead{$P$} & \colhead{$H_V$}  & \colhead{Group} & \colhead{Subgroup} \\
  \colhead{Designation} & \colhead{(au)} & \colhead{($10^6$~km)} & \colhead{}  & \colhead{(deg)}  & \colhead{(days)} & \colhead{}} 
\startdata
S/2007 S 8  & 0.1140 & 17.05  &  0.49  &  36.2  &  836.9  &  15.97 & Gallic & Albiorix \\
Albiorix (26)  & 0.1092 & 16.33  &  0.48  &  36.8  &  783.5  &  11.17 & Gallic & Albiorix \\
Erriapus (28)  & 0.1170 & 17.51  &  0.476  &  37.1  &  871.1  &  13.71 & Gallic & Albiorix \\
S/2004 S 24  & 0.1560 & 23.34  &  0.071  &  37.4  &  1341.3  &  15.98 & Gallic & - \\
Tarvos (21)  & 0.1218 & 18.22  &  0.522  &  37.8  &  926.4  &  13.04 & Gallic & Albiorix \\
Bebhionn (37)  & 0.1138 & 17.03  &  0.459  &  38.5  &  834.9  &  14.99 & Gallic & Albiorix \\
Sat LX & 0.1140 & 17.06  &  0.485  &  38.6  &  837.8  &  15.83 & Gallic & Albiorix \\
S/2006 S 12  & 0.1308 & 19.57  &  0.542  &  38.6  &  1035.1  &  16.2 & Gallic & Albiorix \\
S/2020 S 4  & 0.1219 & 18.24  &  0.495  &  40.1  &  926.9  &  17.01 & Gallic & Albiorix \\
S/2020 S 3  & 0.1207 & 18.06  &  0.142  &  46.0  &  908.2  &  16.38 & Inuit & Siarnaq \\
S/2019 S 14  & 0.1193 & 17.85  &  0.172  &  46.2  &  893.1  &  16.32 & Inuit & Siarnaq \\
S/2019 S 6  & 0.1217 & 18.21 &  0.120  &  46.4  &  919.7  &  15.73 & Inuit & Siarnaq \\
Siarnaq (29)  & 0.1195 & 17.88  &  0.309  &  47.8  &  895.6  &  10.61 & Inuit & Siarnaq \\
Kiviuq (24)  & 0.0756 & 11.31  &  0.275  &  48.0  &  449.1  &  12.67 & Inuit & Kiviuq \\
S/2005 S 4  & 0.0757 & 11.32  &  0.315  &  48  &  450.2  &  15.69 & Inuit & Kiviuq \\
S/2004 S 31  & 0.1170 & 17.50  &  0.159  &  48.1  &  866.1  &  15.63 & Inuit & Siarnaq \\
S/2020 S 1  & 0.0758 & 11.34  &  0.337  &  48.2  &  451.1  &  15.92 & Inuit & Kiviuq \\
S/2020 S 5  & 0.1229 & 18.39  &  0.22  &  48.2  &  933.9  &  16.59 & Inuit & Siarnaq \\
Paaliaq (20)  & 0.1003 & 15.00  &  0.378  &  48.5  &  687.0  &  11.71 & Inuit & - \\
Tarqeq (52)  & 0.1186 & 17.75  &  0.143  &  48.7  &  885.0  &  14.82 & Inuit & Siarnaq \\
Ijiraq (22)  & 0.0758 & 11.34  &  0.293  &  49.2  &  451.4  &  13.27 & Inuit & Kiviuq \\
S/2019 S 1  & 0.0752 & 11.25  &  0.384  &  49.5  &  445.5  &  15.32 & Inuit & Kiviuq \\
Bestla (39)  & 0.1360 & 20.34  &  0.486  &  138.3  &  1087.2  &  14.61 & Norse & - \\
Narvi (31)  & 0.1289 & 19.29  &  0.441  &  142.2  &  1003.9  &  14.52 & Norse & - \\
S/2019 S 11  & 0.1381 & 20.66  &  0.513  &  144.6  &  1115.0  &  16.25 & Norse & - \\
Hyrrokkin (44)  & 0.1226 & 18.34  &  0.336  &  149.9  &  931.9  &  14.34 & Norse & - \\
Skathi (27)  & 0.1041 & 15.58  &  0.281  &  151.6  &  728.1  &  14.41 & Norse & - \\
S/2019 S 19  & 0.1540 & 23.04  &  0.458  &  151.8  &  1318.0  &  16.51 & Norse & Kari \\
Kari (45)  & 0.1473 & 22.03  &  0.469  &  153.0  &  1231.2  &  14.49 & Norse & Kari \\
S/2004 S 21  & 0.1548 & 23.16  &  0.394  &  153.2  &  1328.6  &  16.21 & Norse & Kari \\
S/2004 S 36  & 0.1564 & 23.39  &  0.625  &  153.3  &  1349.4 &  16.11 & Norse & Kari \\
S/2004 S 45  & 0.1316 & 19.69  &  0.551  &  154  &  1038.7  &  15.97 & Norse & Kari \\
Geirrod (66)  & 0.1488 & 22.26  &  0.539  &  154.3  &  1251.1  &  15.89 & Norse & Kari \\
S/2019 S 18  & 0.1547 & 23.14  &  0.509  &  154.6  &  1327.1  &  16.56 & Norse & Kari \\
S/2019 S 17  & 0.1519 & 22.72  &  0.546  &  155.5  &  1291.4  &  15.86 & Norse & Kari \\
S/2006 S 1  & 0.1253 & 18.75  &  0.105  &  156.0  &  964.2  &  15.65 & Norse & Kari \\
S/2006 S 3  & 0.1427 & 21.35  &  0.432  &  156.1  &  1174.8  &  15.65 & Norse & Kari \\
S/2019 S 20  & 0.1583 & 23.68  &  0.354  &  156.1  &  1375.5  &  16.73 & Norse & Kari \\
Farbauti (40)  & 0.1356 & 20.29  &  0.249  &  156.2  &  1087.3  &  15.75 & Norse & Kari \\
S/2019 S 15  & 0.1416 & 21.19  &  0.257  &  157.8  &  1161.6  &  16.59 & Norse & Mundilfari \\
S/2004 S 37  & 0.1066 & 15.94  &  0.448  &  158.2  &  755.6  &  15.92 & Norse & Mundilfari \\
S/2007 S 5  & 0.1059 & 15.84  &  0.104  &  158.4  &  746.9  &  16.23 & Norse & Mundilfari \\
Bergelmir (38)  & 0.1288 & 19.27  &  0.145  &  158.8  &  1005.5  &  15.16 & Norse & Mundilfari \\
Thiazzi (63) & 0.1576 & 23.58  &  0.512  &  158.8  &  1366.7  &  15.91 & Norse & Mundilfari \\
S/2019 S 5  & 0.1276 & 19.09  &  0.216  &  158.8  &  990.4  &  16.65 & Norse & Mundilfari \\
Beli (61)  & 0.1384 & 20.70  &  0.087  &  158.9  &  1121.7  &  16.09 & Norse & Mundilfari \\
S/2007 S 9  & 0.1348 & 20.17  &  0.360  &  158.9  &  1078.1  &  16.06 & Norse & Mundilfari \\
Skoll (47)  & 0.1178 & 17.62  &  0.463  &  159.4  &  878.4  &  15.41 & Norse & Mundilfari \\
S/2019 S 9  & 0.1361 & 20.36  &  0.433  &  159.5  &  1093.1  &  16.27 & Norse & Mundilfari \\
S/2004 S 49  & 0.1497 & 22.40  &  0.453  &  159.7  &  1264.3  &  15.97 & Norse & Mundilfari \\
Gunnlod (62)  & 0.1413 & 21.14  &  0.251  &  160.3  &  1158.0  &  15.57 & Norse & Mundilfari \\
S/2004 S 47  & 0.1073 & 16.05  &  0.291  &  160.9  &  762.5  &  16.29 & Norse & Mundilfari \\
S/2006 S 15  & 0.1457 & 21.80  &  0.117  &  161.1  &  1214.0  &  16.22 & Norse & Mundilfari \\
S/2020 S 7  & 0.1163 & 17.40  &  0.500  &  161.4  &  861.2  &  16.79 & Norse & Mundilfari \\
S/2020 S 9  & 0.1699 & 25.41  &  0.531  &  161.4  &  1532.6  &  16.02 & Norse & Mundilfari \\
S/2006 S 10  & 0.1269 & 18.98  &  0.151  &  161.6  &  983.1  &  16.43 & Norse & Mundilfari \\
S/2020 S 8  & 0.1469 & 21.97  &  0.252  &  161.8  &  1228.1  &  16.41 & Norse & Mundilfari \\
\enddata
\end{deluxetable*}

\begin{deluxetable*}{lcccccccc}
  \tabletypesize{\scriptsize}
  \setlength{\tabcolsep}{0.1in}
  \tablecaption{Continuation of \autoref{tab:orbit}. \label{tab:orbit2}}
  \tablewidth{0pt}
  \tablehead{ \colhead{Name/} & \multicolumn{2}{c}{$a$} & \colhead{$e$} & \colhead{$i$}  & \colhead{$P$} & \colhead{$H_V$}  & \colhead{Group} & \colhead{Subgroup} \\
  \colhead{Designation} & \colhead{(au)} & \colhead{($10^6$~km)} & \colhead{}  & \colhead{(deg)}  & \colhead{(days)} & \colhead{}} 
\startdata
S/2004 S 48  & 0.1480 & 22.14  &  0.374  &  161.9  &  1242.4  &  15.95 & Norse & Mundilfari \\
S/2006 S 13  & 0.1334 & 19.95  &  0.313  &  162.0  &  1060.6  &  16.05 & Norse & Mundilfari \\
S/2019 S 16  & 0.1556 & 23.27  &  0.250  &  162.0  &  1340.9  &  16.68 & Norse & Mundilfari \\
S/2004 S 53  & 0.1556 & 23.28  &  0.240  &  162.6  &  1342.4  &  16.16 & Norse & Mundilfari \\
Jarnsaxa (50)  & 0.1288 & 19.27  &  0.218  &  163.0  &  1006.5  &  15.62 & Norse & Mundilfari \\
Gridr (54)  &  0.1287 & 19.25  &  0.187  &  163.9  &  1004.7  &  15.77 & Norse & Mundilfari \\
S/2019 S 10  & 0.1384 & 20.70  &  0.248  &  163.9  &  1123.0  &  16.66 & Norse & Mundilfari \\
S/2004 S 50  & 0.1494 & 22.35  &  0.45  &  164.0  &  1260.4  &  16.4 & Norse & Mundilfari \\
S/2006 S 16  & 0.1452 & 21.72  &  0.204  &  164.1  &  1207.5  &  16.54 & Norse & Mundilfari \\
Fenrir (41)  & 0.1493 & 22.33  &  0.137  &  164.5  &  1260.2  &  15.89 & Norse & Mundilfari \\
S/2004 S 12  & 0.1324 & 19.80  &  0.337  &  164.7  &  1048.6  &  15.91 & Norse & Mundilfari \\
S/2004 S 7  & 0.1426 & 21.33  &  0.511  &  164.9  &  1173.9  &  15.56 & Norse & Mundilfari \\
Eggther (59)  & 0.1326 & 19.84  &  0.157  &  165.0  &  1052.3  &  15.39 & Norse & Mundilfari \\
S/2004 S 52  & 0.1768 & 26.45  &  0.292  &  165.3  &  1633.9  &  16.5 & Norse & Mundilfari \\
Hati (43)  & 0.1317 & 19.70  &  0.372  &  165.4  &  1040.1  &  15.45 & Norse & Mundilfari \\
S/2020 S 10  & 0.1692 & 25.31  &  0.296  &  165.6  &  1527.2  &  16.86 & Norse & Mundilfari \\
S/2004 S 41  & 0.1210 & 18.10  &  0.301  &  165.7  &  914.6  &  16.31 & Norse & Mundilfari \\
S/2004 S 42  & 0.1219 & 18.24  &  0.157  &  165.7  &  925.9  &  16.11 & Norse & Mundilfari \\
S/2004 S 39  & 0.1550 & 23.19  &  0.101  &  165.9  &  1335.9  &  16.14 & Norse & Mundilfari \\
Aegir (36)  & 0.1381 & 20.66  &  0.255  &  166.1  &  1119.3  &  15.51 & Norse & Mundilfari \\
S/2007 S 6  & 0.1239 & 18.54  &  0.169  &  166.5  &  949.5  &  16.36 & Norse & Mundilfari \\
S/2006 S 14  & 0.1408 & 21.06  &  0.060  &  166.7  &  1152.7  &  16.5 & Norse & Mundilfari \\
S/2019 S 3  & 0.1142 & 17.08  &  0.249  &  166.9  &  837.7  &  16.22 & Norse &  Mundilfari\\
S/2020 S 6  & 0.1420 & 21.25  &  0.480  &  166.9  &  1167.9  &  16.55 & Norse & Mundilfari \\
Mundilfari (25)  & 0.1243 & 18.59  &  0.212  &  166.9  &  952.9  &  14.57 & Norse & Mundilfari\\
S/2019 S 12  & 0.1396 & 20.89  &  0.475  &  167.1  &  1138.9  &  16.33 & Norse & Mundilfari \\
S/2004 S 44  & 0.1305 & 19.52  &  0.129  &  167.7  &  1026.2  &  15.82 & Norse & Mundilfari \\
S/2004 S 28  & 0.1462 & 21.87  &  0.159  &  167.9  &  1220.7  &  15.77 & Norse & Mundilfari\\
S/2004 S 17  & 0.1317 & 19.70  &  0.162  &  167.9  &  1040.9  &  15.95 & Norse & Mundilfari\\
Loge (46)  & 0.1532 & 22.92  &  0.192  &  166.9  &  1311.8  &  15.36 & Norse & Mundilfari \\
Sat LXIV  & 0.1614 & 24.14  &  0.280  &  168.3  &  1420.8  &  16.15 & Norse & Mundilfari\\
Surtur (48)  & 0.1521 & 22.75  &  0.448  &  168.4  &  1295.6  &  15.77 & Norse & Mundilfari\\
S/2006 S 17  & 0.1496 & 22.38  &  0.425  &  168.7  &  1264.5  &  16.01 & Norse & Mundilfari\\
S/2004 S 13  & 0.1233 & 18.45  &  0.265  &  169.0  &  942.6  &  16.25 & Norse & Mundilfari\\
S/2004 S 40  & 0.1075 & 16.08  &  0.297  &  169.2  &  764.6  &  16.28 & Norse & Mundilfari\\
S/2007 S 7  & 0.1065 & 15.93  &  0.217  &  169.2  &  754.3  &  16.24 & Norse & Mundilfari\\
S/2005 S 5  & 0.1428 & 21.37  &  0.588  &  169.5  &  1177.8  &  16.36 & Norse & Mundilfari\\
S/2006 S 18  & 0.1521 & 22.76  &  0.131  &  169.5  &  1298.4  &  16.1 & Norse & Mundilfari\\
Fornjot (42)  & 0.1667 & 24.94  &  0.213  &  170.0  &  1494.1  &  15.12 & Norse & Mundilfari\\
S/2019 S 4  & 0.1200 & 17.95  &  0.408  &  170.1  &  903.9  &  16.46 & Norse & Mundilfari\\
S/2020 S 2  & 0.1195 & 17.87  &  0.152  &  170.7  &  897.6  &  16.89 & Norse & Mundilfari\\
S/2004 S 43  & 0.1266 & 18.94  &  0.432  &  171.1  &  980.1  &  16.34 & Norse & Mundilfari\\
S/2004 S 51  & 0.1685 & 25.21  &  0.201  &  171.2  &  1519.4  &  16.13 & Norse & Mundilfari\\
S/2019 S 21  & 0.1767 & 26.44  &  0.155  &  171.9  &  1636.4  &  16.18 & Norse & Mundilfari\\
Ymir (19)  & 0.1534 & 22.95  &  0.338  &  172.3  &  1315.1  &  12.41 & Norse & Phoebe \\
S/2019 S 8  & 0.1356 & 20.29  &  0.311  &  172.8  &  1088.9  &  16.28 & Norse & Phoebe \\
Sat LVIII & 0.1745 & 26.10  &  0.147  &  173.0  &  1604.0  &  15.7 & Norse & Phoebe \\
S/2006 S 9  & 0.0963 & 14.41  &  0.248  &  173.0  &  647.9  &  16.48 & Norse & Phoebe \\
S/2006 S 20  & 0.0882 & 13.19  &  0.206  &  173.1  &  567.3  &  15.75 & Norse & Phoebe \\
S/2019 S 2  & 0.1107 & 16.56  &  0.279  &  173.3  &  799.8  &  16.49 & Norse & Phoebe \\
S/2007 S 3  & 0.1311 & 19.61  &  0.150  &  173.8  &  1034.4  &  15.74 & Norse & Phoebe \\
S/2007 S 2  & 0.1066 & 15.94  &  0.232  &  174.0  &  754.9  &  15.59 & Norse & Phoebe \\
S/2006 S 11  & 0.1318 & 19.71  &  0.143  &  174.1  &  1042.3  &  16.47 & Norse & Phoebe \\
Greip (51)  & 0.1229 & 18.38  &  0.317  &  174.1  &  937.0  &  15.33 & Norse & Phoebe \\
S/2019 S 7  & 0.1349 & 20.18  &  0.232  &  174.2  &  1080.6  &  16.29 & Norse & Phoebe \\
Gerd (57)  & 0.1400 & 20.95  &  0.518  &  174.4  &  1143.0  &  15.87 & Norse & Phoebe \\
Thrymr (30)  & 0.1359 & 20.33  &  0.467  &  175.0  &  1092.2  &  14.33 & Norse & Phoebe \\
Phoebe (9)  & 0.0864 & 12.93  &  0.164  &  175.2  &  550.3  &  6.73 & Norse & Phoebe \\
S/2006 S 19  & 0.1591 & 23.80  &  0.467  &  175.5  &  1389.3  &  16.07 & Norse & Phoebe \\
Skrymir (56)  & 0.1434 & 21.45  &  0.437  &  175.6  &  1185.1  &  15.62 & Norse & Phoebe \\
Suttungr (23)  & 0.1296 & 19.39  &  0.116  &  175.7  &  1016.7  &  14.55 & Norse & Phoebe \\
S/2004 S 46  & 0.1371 & 20.51  &  0.249  &  177.2  &  1107.6  &  16.4 & Norse & Phoebe \\
S/2019 S 13  & 0.1401 & 20.96  &  0.318  &  177.3  &  1144.8  &  16.68 & Norse & Phoebe \\
Alvaldi (65)  & 0.1470 & 21.99  &  0.238  &  177.4  &  1232.2  &  15.62 & Norse & Phoebe \\
Angrboda (55)  & 0.1376 & 20.59  &  0.216  &  177.7  &  1114.1  &  16.17 & Norse & Phoebe \\
\enddata
\end{deluxetable*}

\subsection{Gallic Group}
\label{sec:gallic}

\begin{figure}[ht]
\plotone{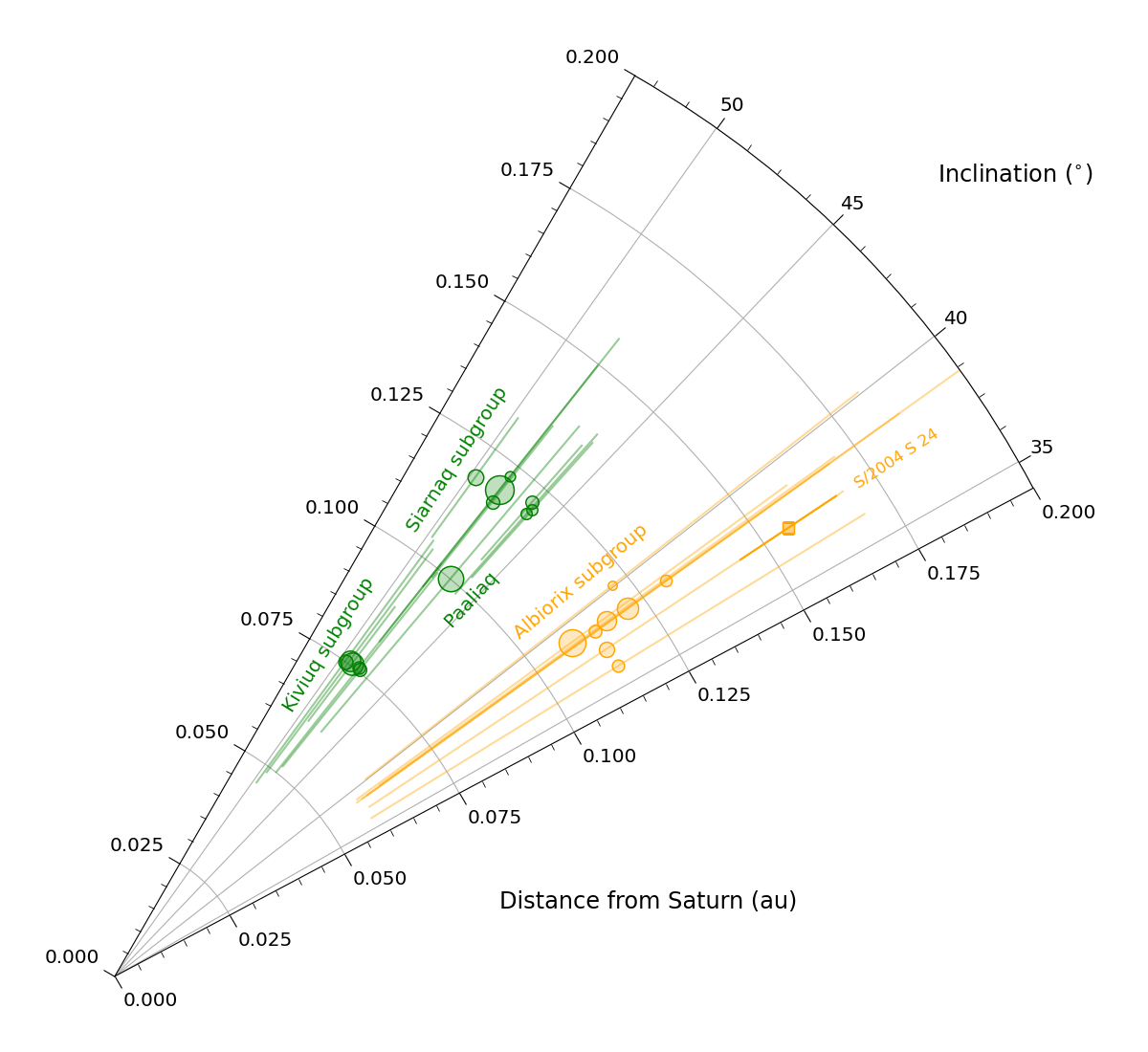}
\caption{Rose diagram of the irregular moons of Saturn with direct orbits. Each circle represents the time averaged inclination and semi-major axis of a moon, with the size of the circle indicating the relative size of the moon. Additionally, each moon has a line that goes from the time-averaged pericentre to the time-averaged apocentre. Yellow represents moons that belong to the Gallic group, and green represents the Inuit group. 
The gap between S/2004 S 24 (represented by a square instead of a circle) and the rest of the Gallic group has decreased with the inclusion of the new moons, but still remains large (see \autoref{sec:gallic} for discussion).
}
\label{fig:directrose}
\end{figure}

The Gallic group only gained three new members, which makes it now 9 moons in total (see \autoref{fig:directrose}).
Of the six previously known Gallic group members, all but S/2004 S 24 have similar $a$. This clustering, which we will refer to as the Albiorix subgroup, also includes all three of the new members.
Interestingly, Albiorix has the lowest $a$ of all known Gallic members. Perhaps the Albiorix subgroup, or even the Gallic group as a whole, was created 
via a cratering event on the leading hemisphere of Albiorix, with this still (mostly) intact parent body remaining close to its pre-collision orbit.
If the crater location was near the apex point of Albiorix's velocity, all the fragments would be ejected in the direction of the leading hemisphere; their relative $\Delta v$ boost would have some component in the direction of motion of the parent body, producing larger $a$ (see \autoref{sec:colgal}).
Such a leading-hemisphere impact event is greatly favoured when the projectile is moving in the opposite orbital sense from the target, and naturally results in a higher relative impact speed (about twice the orbital speed for a direct encounter), allowing retrograde projectiles to provide a much larger kinetic energy to the impact.
Given that there are significantly more retrograde irregular moons compared to direct ones, a leading-hemisphere collision is a plausible scenario.

One of the new Gallic moons, S/2006 S 12, has reduced the size of the semi-major axis gap between the Albiorix subgroup and S/2004 S 24 but this gap still remains larger
than the pericentre of S/2004 S 24.  
The disconnect of S/2004 S 24 from the rest of the group, in terms of having a significantly larger $a$ and being the only Gallic member whose perihelion is not lower than the
semi-major axis of Albiorix itself, means that the question still remains whether S/2004 S 24 comes from the same parent body as the rest of the Gallic group. We explore this further in \autoref{sec:colgal}.

%------------------------------------ FIGURE 2 -------------------------------
%------------------------------ end Figure 2 --------------------------------------

\subsection{Inuit Group}

The Inuit group has more than doubled their number of known members, from 6 to 13. With the addition of these new Inuit moons, some structure within the group has become clear (see \autoref{fig:directrose}). All of the new Inuit moons have semi-major axes close to that of a previously known moon, producing two subgroups which were both %briefly [Mike didn't like the briefly]
mentioned by \citet{Sheppard2023}.

Three of the new Inuit moons, S/2019 S 1, S/2020 S 1, and S/2005 S 4, have very similar $a$ to Kiviuq and Ijiraq, creating a subgroup within the Inuit group.
The pair of Kiviuq and Ijiraq has long been thought to have a common origin \citep{Turrini2008,Denk2018}.
Thus, with the addition of the new moons, it has become clear that the Kiviuq subgroup (named after the largest member) is highly likely a collisional family with very low velocity dispersion.

There is a similar case with the moons surrounding Siarnaq. When there was only Tarqeq with similar $a$ and $i$ to Siarnaq, \citet{Denk2018} labelled the two as a potential co-orbiting pair that separated. Since then S/2004 S 31 and now the new moons of S/2019 S 6, S/2020 S 3, S/2020 S 5, and S/2019 S 14 have been revealed to have orbits similar to that of Siarnaq. Thus, the Siarnaq subgroup, consisting of 7 members, is likely another collisional family in the Inuit group.

Interestingly, unlike the rest of the Inuit members, Paaliaq currently does not have any companions that share similar $a$. Perhaps Paaliaq, with an $a$ that is about halfway between the Kiviuq and Siarnaq subgroups, is part of a cluster, but the other members are too small to be detected currently.

The big question about the Inuit group still remains: Do all members share a common origin? 
One scenario is that the Inuit group originated from a single moon that has been broken up by multiple collisions. 
Although it is hard to imagine how three distinct clusterings with a large variation in semi-major axes can arise from what was initially a single object.
Another scenario is that the Kiviuq subgroup, the Siarnaq subgroup, and Paaliaq are not related to each other.
However, the probability that three unrelated clusterings share a similar $i$ is very low,
especially since there appears to be only one other direct group in the saturnian system.
Perhaps they initially had different inclinations and over time they became aligned, although such a mechanism is currently unknown.

\subsection{Norse Group}

\citet{Gladman2001} initially suggested that Skathi (S/2000 S 8) should be kept separate from
the rest of the so-called Phoebe subgroup due to its significantly different orbital
inclination; in the end, all retrograde saturnians have been receiving norse-themed names.
Even the postulated `Phoebe group' was remarked to be unusual, in that Phoebe is very much larger than the other 
4 discoveries from 2000: Ymir, Thrymr, Suttungr (which have Phoebe-like inclinations but much larger semi-major axes) and Mundilfari, which has an inclination offset from Phoebe much larger than the spread seen in the two direct inclination groups).

\begin{figure}[ht]
\plotone{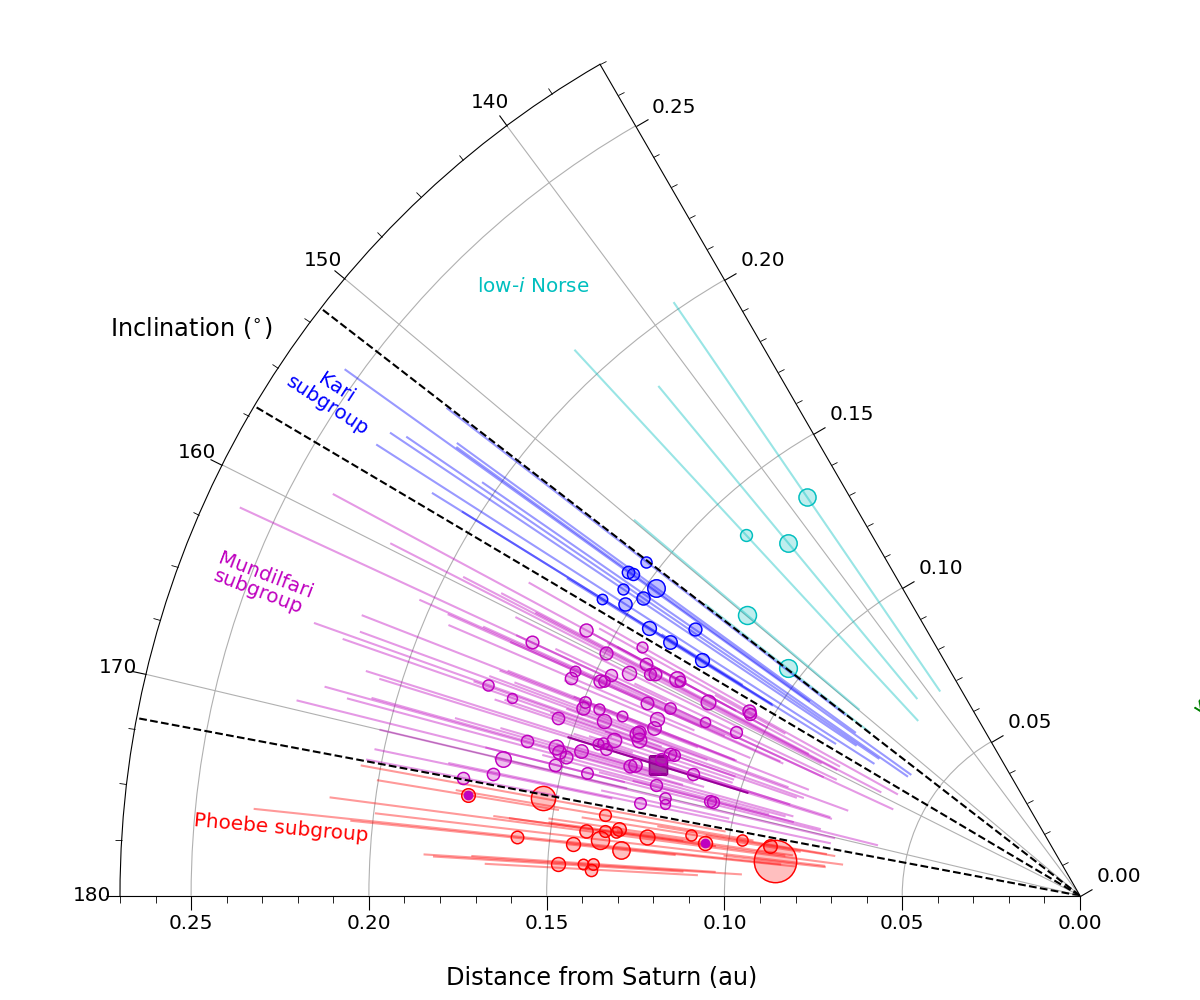}
\caption{Rose diagram of the retrograde saturnian irregulars. See Fig.~\ref{fig:directrose}'s caption.
Here, red represents moons belonging to the Phoebe subgroup, magenta represents the Mundilfari subgroup, blue represents the Kari subgroup, and cyan represents low-i ($<$151$^\circ$) Norse members. Mundilfari is represented by a square and is a darker shade of purple, in order to distinguish the moon from the rest of the subgroup named after it. Two Phoebe subgroup members that may be outlying members of the Mundilfari subgroup are indicated with an additional inset magenta dot.}
\label{fig:retrorose}
\end{figure}

\begin{figure}[ht]
\plotone{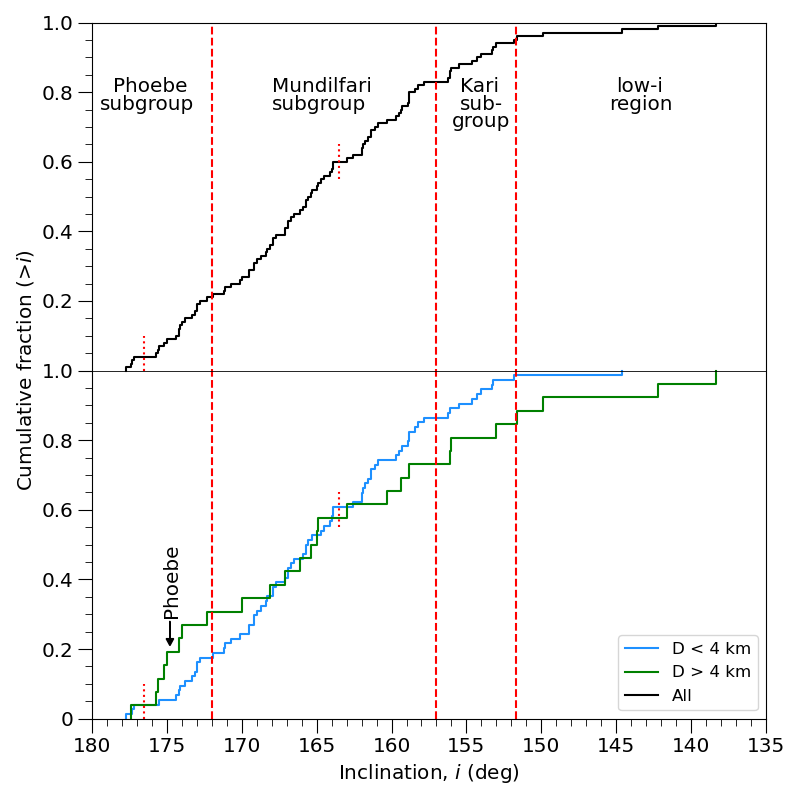}
\caption{Inverse cumulative inclination distribution 
of the full Norse/retrograde population (top panel) and then split up 
into large ($>4$~km) and small ($<4$~km) members (bottom panel). 
We divide the Norse group up into three sub-groups with red dashed lines marking the boundaries. There are two sets of short red dotted lines in each panel, one at 176.5$^{\circ}$ indicating a potential further subdivision within the Phoebe subgroup
(see \autoref{sec:Phoebe}) 
and one at 163$^{\circ}$ indicating a minor dearth of moons within the Mundilfari subgroup (see \autoref{sec:Munsplit}).}
\label{fig:icumu}
\end{figure}

Like the previously known moons, the vast majority of the new moons belong to the Norse/retrograde group. 
Of the 64 new moons, 54 are Norse group members. This is a slightly larger fraction than that of previously known moons (46/58). 
The retrograde population has now exploded in number into a very complicated distribution that is far more difficult to interpret than the tightly-clustered direct groups.
At first glance, the rose plot of the retrograde moons (\autoref{fig:retrorose}) does not cleanly manifest much structure in the orbital distribution. 
Looking at a cumulative inclination distribution, however, it 
appears that the Norse group can be broken up by using inclination cuts 
tied to subtle features (top panel of \autoref{fig:icumu}).

%------------------------------------ FIGURE 3 -------------------------------
%------------------------------ end Figure 3 --------------------------------------

%------------------------------------ FIGURE 4 -------------------------------
%------------------------------ end Figure 4 --------------------------------------

\subsubsection{Phoebe Subgroup}
\label{sec:Phoebe}

\citet{Ashton2021} attempted to subdivide the Norse group by defining the Phoebe subgroup as moons with inclinations within 3$^{\circ}$ of Phoebe's value. 
From the top panel of \autoref{fig:icumu} a clear absence of moons (relative to surrounding inclinations) can be seen at 172$^{\circ}$, suggesting there is a division between Phoebe-like inclinations and the rest. This division can be seen more clearly when looking at just the large (D $>$ 4~km) Norse members (bottom panel of \autoref{fig:icumu}).
Here we will use the criteria $i>172^{\circ}$ to define the Phoebe subgroup (although, currently, we get the same Phoebe subgroup members if we use the criteria from \citet{Ashton2021}).
The recent batch of new discoveries has added 10 more members to the Phoebe subgroup for a total of 21 (see \autoref{tab:orbit} for the full list).

We caution, however, that the saturnocentric distance variation of a large fraction of the Phoebe subgroup members do not overlap with that of Phoebe (i.e. the pericenter of some members is larger than the apocenter of Phoebe).
As such, a scenario of the subgroup being created by a single collision off Phoebe seems unlikely.
There is a small clustering in Phoebe subgroup members around 20$\times 10^6$~km$\simeq$0.13~au which may be a good candidate for a small
collisional subfamily.
Furthermore, there are four moons, Angrboda, Alvaldi, S/2019 S 13, and S/2004 S 46, which have similar $a \sim0.14$~au and $i$, 177.2--177.7$^{\circ}$.
This close clustering in $i$, and the fact that they are 1.6$^{\circ}$ away from the rest of the Phoebe subgroup (see \autoref{fig:icumu}), could be an indication of another small collisional subfamily. 
Curiously, the large irregular Ymir is still quite isolated at $a\simeq 0.15$~au and $i\simeq172^\circ$.
Lastly, the lowest-$a$ members of the Phoebe subgroup, S/2006 S 9 and S/2006 S 20 (maybe even S/2019 S 2 and S/2007 S 2 too), could be fragments of a cratering event off Phoebe. %Slightly tweaked it

\subsubsection{Kari Subgroup and low inclination Norse moons}

At the low-$i$ end of the Norse group (closest to $i=90^\circ$) the cumulative distribution (top panel of \autoref{fig:icumu}) flattens out when going to smaller $i$. 
This flattening occurs at two $i$'s, 157$^{\circ}$ and 152.5$^{\circ}$.
This suggests that these $i$'s might be good locations for having subgroup boundaries.
We decided to extend the lower limit to 151.7$^{\circ}$ so that this subgroup includes S/2019 S 19, which has a $a$ similar to that of a majority of the subgroup members. %Forced change due to updated orbits
%We could easily put the lower dividing line at $i$ = 152$^{\circ}$, but since S/2019 S 19 has an $a$ that is similar to a majority of the members in the Kari subgroup we suggest it should be included in the subgroup. 
As such, we define the Kari subgroup (named after the largest member) as the moons with 151.7$^{\circ} <$ $i <$ 157$^{\circ}$.
The tightly packed nature of the Kari subgroup, without the 3-4 lowest $a$ members, makes it a good candidate for a collisional family.

Of the five Norse moons with the lowest $i$, which were the only moons not placed in any subgroups, Skathi and Hyrrokkin could potentially be related (although there is a large separation between their $a$'s). %Forced change due to updated orbits
Narvi, S/2019 S 11, and Bestla have similar $a$'s but are a bit spread in $i$.
We note that this population is approaching $i=90^\circ$ closely enough that the Kozai effect may begin to operate, like that observed for three Inuit moons by \citet{Jacobson2022}, with the potential for long-term instability of the moons \citet{Gladman2001} setting a boundary for this group.

\subsubsection{Mundilfari Subgroup}
\label{sec:core}

The remaining Norse moons (with 157$^{\circ} > i > 172^{\circ}$) are put into what we denote the Mundilfari subgroup.
This subgroup could actually be broken into two as there is a dearth of moons around $i$ of 163$^{\circ}$ (top panel of \autoref{fig:icumu}).
However, both regions of the Mundilfari subgroup that are divided by the $i$ = 163$^{\circ}$ line have significantly higher fractions of small (D $<$ 4~km) Norse moons compared with the fractions of large (D $>$ 4~km) Norse moons (bottom panel of \autoref{fig:icumu}). 
All the other inclination ranges either have fractions of large moons that are similar or larger
than that of small moons (currently).
This suggests that both of these two $i$ ranges have significantly steeper size distributions compared to the other saturnian irregulars. 
As such, we have kept it as one subgroup. 
Further discussion of size distributions is presented in \autoref{sec:sizedist}.

Using the inclination distribution is a good method to define Norse subgroups as collisional fragments will clump together in $i$.
Although, having just a single $i$ as the boundary between groups will most likely result in moons being put into sub-groups that they do not belong to.
Two good examples of this are discussed in \autoref{sec:sizedist}. 

\section{Size Distribution}
\label{sec:sizedist}

\begin{figure}[ht]
\plotone{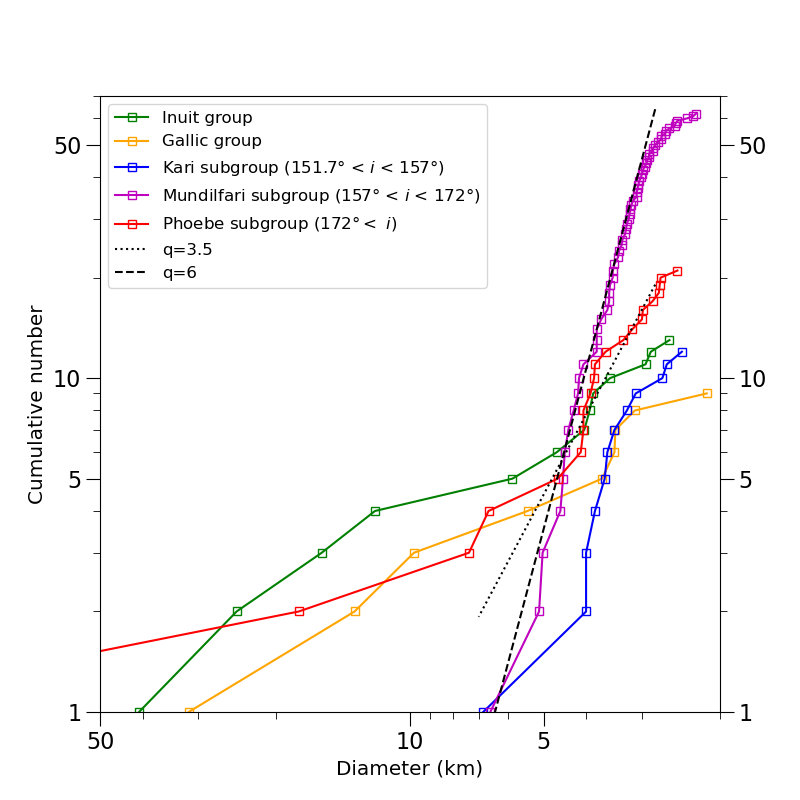}
\caption{Size distribution of different groups/subgroups in the saturnian irregular moon population. 
The method used to calculate diameters is described in \autoref{sec:size}.
Two lines have been added for reference, one with a collisional equilibrium, q = 3.5 (dotted line), and one with a similar slope to that of the Mundilfari subgroup, q = 6 (dashed line). Note: the five $i <$ 151$^{\circ}$ Norse group members are not represented in this plot.}
\label{fig:sizedist}
\end{figure}

Although the inclination study was done first, and the boundaries of these divisions are obviously not very strong, we later discovered that the splits as we made them resulted in a dramatically different size distribution for the Mundilfari subgroup when compared to all other saturnians.
When one splits up the saturnian irregular size distribution into different groups/subgroups (\autoref{fig:sizedist})
one can see that the Mundilfari subgroup is significantly steeper than the other clusters over a large range of diameters.
As shown by one of the reference slopes (dotted line) in \autoref{fig:sizedist} the Mundilfari subgroup has a fairly consistent differential power law index, $q$, of approximately 6 down until at least $D \sim$ 3~km, at which point incompleteness causes the distribution to flatten out.

%------------------------------------ FIGURE 4 -------------------------------
%------------------------------ end Figure 4 ----------------------------------

The other two Norse subgroups, Phoebe and Kari, generally have much less steep slopes in the same size range (3--6~km), similar to that of collisional equilibrium, $q = 3.5$ \citep{Dohnanyi1969}. 
Even shallower still are the two direct groups, Gallic and Inuit, which both have shallow slopes of $q \sim$ 2 throughout.

Both the Phoebe and Kari subgroups have a narrow size range around 4~km where the size distributions have similar steepness to that of the Mundilfari subgroup. 
As mentioned in \autoref{sec:core}, there is likely cross-contamination between what we have defined as subgroups. In the size range where the Phoebe subgroup has a very steep slope (3.7--4.7~km), 4 out of the 8 members have $i$ within 2$^{\circ}$ of the Mundilfari subgroup (i.e. 172$^{\circ}$ $< i <$ 174$^{\circ}$). One of the four moons is S/2006 S 20 which, as mentioned in \autoref{sec:Phoebe} is likely a collisional fragment off of Phoebe. 
Another is S/2007 S 3, with an $a$ of 0.1311~au, is at the outer edge of a large cluster within the Phoebe subgroup, and is thus more likely to be associated with this cluster and not to the Mundilfari subgroup.
The other two moons, S/2007 S 2 and Sat LVIII, could easily belong to the Mundilfari subgroup, as they are fairly isolated in $a$-$i$ (see \autoref{fig:retrorose}).
Additionally, Sat LVIII has a similar $a$ to Mundilfari members just on the other side of the inclination divide. If both of these moons are indeed members of the Mundilfari subgroup and not the Phoebe subgroup, the slope of the size distribution at around 4~km would be significantly shallower for the Phoebe subgroup.
A similar argument could be made for the Kari subgroup. After Kari, the next three largest members, S/2006 S 3, S/2006 S 1, and Farbauti, are all within 1$^{\circ}$ of the Mundilfari boundary and all have $a$'s lower than the main Kari subgroup cluster.
Additionally, small-number statistics might play a part.
%More Kari members need to be found to get a better understanding of the sub-group's size distribution.

\citet{Ashton2021} found that the saturnian irregulars, as a whole, had $q=4.9$ from $D=3$ to 4~km and postulated that this steep slope was due to a recent collision (in the last Gyr). 
Catastrophic collisions (and cratering events) are known to produce steep size distrbutions (see \citet{Durda2007} for example).
They were unable to determine which members were likely involved in the collision
as almost all of the moons that they detected in the size range did not have known orbits at the time.
However, they did suggest that it was likely located 
in the retrograde population.
The new saturnian irregulars discovered show that the Mundilfari subgroup 
alone is essentially the cause of the steep size distribution found by \citet{Ashton2021}.
Thus, we believe that it is the Mundilfari subgroup that was created by the recent collision hypothesised by \citet{Ashton2021}. 
If this is true then the Mundilfari collisional family is significantly more dispersed in $a$ and $i$ than the other potential collisional families seen among the irregular moons. 

\subsection{Splitting up the Mundilfari Subgroup}
\label{sec:Munsplit}

\begin{figure}[ht]
\plotone{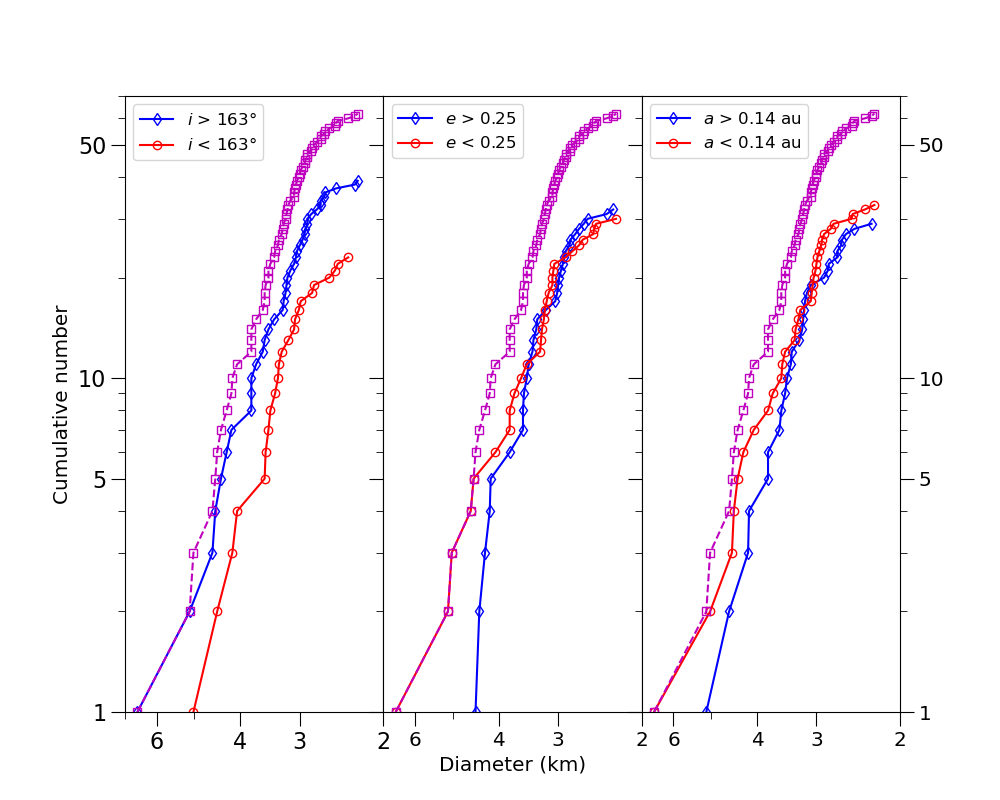}
\caption{A comparison of the size distributions of the Mundilfari subgroup when split into two samples using cuts in inclination (left panel), eccentricity (middle panel), and semi-major axis (right panel). The size distribution of the whole Mundilfari subgroup is shown in magenta in all panels.}
\label{fig:sizedist2}
\end{figure}

We explore the idea that perhaps this steep slope is only present in a subset of the Mundilfari subgroup. If the subgroup is split into two by a potential division at $i$ = 163$^{\circ}$, as discussed in \autoref{sec:core}, the slope of the size distributions of the two regions are very similar to that of the Mundilfari subgroup as a whole (see left panel in \autoref{fig:sizedist2}). Additionally, we split the Mundilfari subgroup into two subsamples two different ways using cuts of $a$ = 0.14~au %($2.1 \times 10^7$~km)
or $e = 0.25$.
The values of the $a$ and $e$ cuts were chosen to split the Mundilfari subgroup into roughly equal numbers.
Like with the $i$ split, the size distribution of the high/low $e$/$a$ all have very high slopes, like the Mundilfari subgroup as a whole (middle and right panels in \autoref{fig:sizedist2}, respectively); the $e$ $<$ 0.25 sample has the shallowest slope with $q \sim 5$.
We therefore believe that the steep slope present in all subdivisions justifies our choice to keep the Mundilfari subgroup intact.

%------------------------------------ FIGURE 5 -------------------------------
%------------------------------ end Figure 5 ----------------------------------

\subsection{Observational Bias}

A caveat that must be taken into consideration is that is sample has not been debaised. 
Even though a majority of saturnian irregulars lie in the two offset fields (Fig.~\ref{fig:tracking}) at any one instant, and the two fields have been observed many times over multiple oppositions, this survey is still somewhat biased towards moons with certain orbits.
The lack of coverage north and south of Saturn (see \autoref{fig:tracking}) results in moons with $i$ closer to 90$^{\circ}$ being harder to detect compared to moons with $i$ closer to 0 or 180$^{\circ}$. This is exemplified by the fact that we never imaged Bestla, the lowest $i$ Norse group member. Another bias, identified in \citet{Ashton2022}, is that the glare from Saturn makes it harder to detect moons with smaller $a$ or pericentre. 
Given these two biases, the Inuit and Gallic subgroups and the low-$i$ Norse moons ($i <$ 151$^{\circ}$) are likely under-represented in the current sample, and thus have size distributions that are likely more steep in reality than what is seen in \autoref{fig:sizedist}.
The Norse subgroups, on the other hand, are less affected by such observational biases,
with the possible exceptions of the low-$a$ part of the Phoebe subgroup and, to a lesser extent, the Kari subgroup.
Therefore, we believe that observational biases in our survey should not influence our analysis or conclusions regarding the Mundilfari subgroup.

\section{Collisional Families Simulations}
\label{sec:col}

Some of the clusters of saturnian irregular moons are tight enough that they could easily be created by a single disruption event (e.g. the Kiviuq subgroup and the Siarnaq subgroup).
On the other hand, the Gallic group and the Mundilfari subgroup have comparatively larger orbital spread, and, as such, we want to test the feasibility of a single collision creating either of these two clusters. 

We calculate the orbital extent of a debris field by generating $n$ ejecta particles by adding a fixed $\Delta {\bf v}$ in random directions to that of the progenitor velocity at the moment of collision, similar to what was done in \citet{Grav2003}. 
To be judged as to realistically reproducing a clustering, we use the criteria that at least one of the generated ejecta must overlap with the secular ranges of $a$, $e$, and $i$ (found in \citet{Jacobson2022}) of every moon in that cluster.
There are aspects which make this current analysis somewhat approximate. Firstly, we did not look at the 3 dimensional $a-e-i$ space, just the $a-e$ and $a-i$ space.
Secondly, when \citet{Jacobson2022} did their analysis, only one of the 64 new moons was announced, so do not have the computed orbital variations. 
Thus, we cannot know whether the ejecta correctly recreate the full range of either cluster.

\subsection{Gallic group}
\label{sec:colgal}

\begin{figure}[ht]
\plottwo{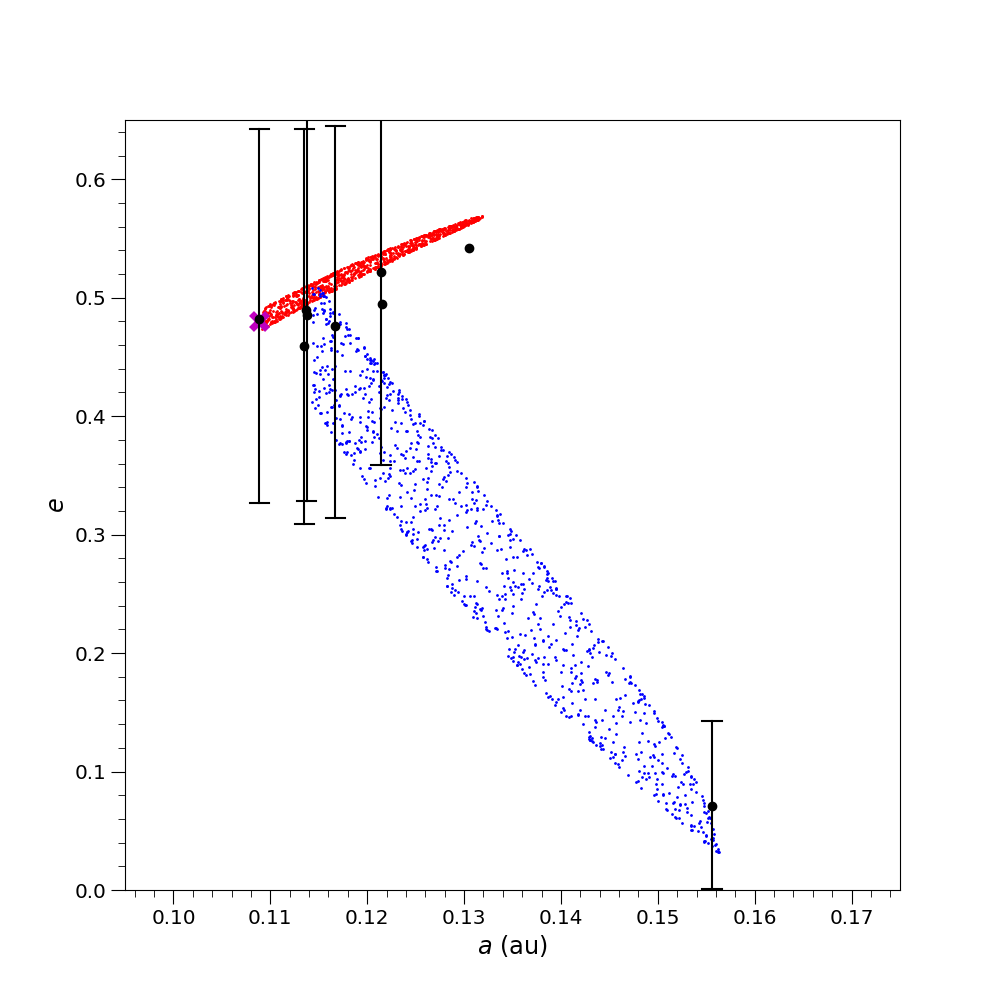}{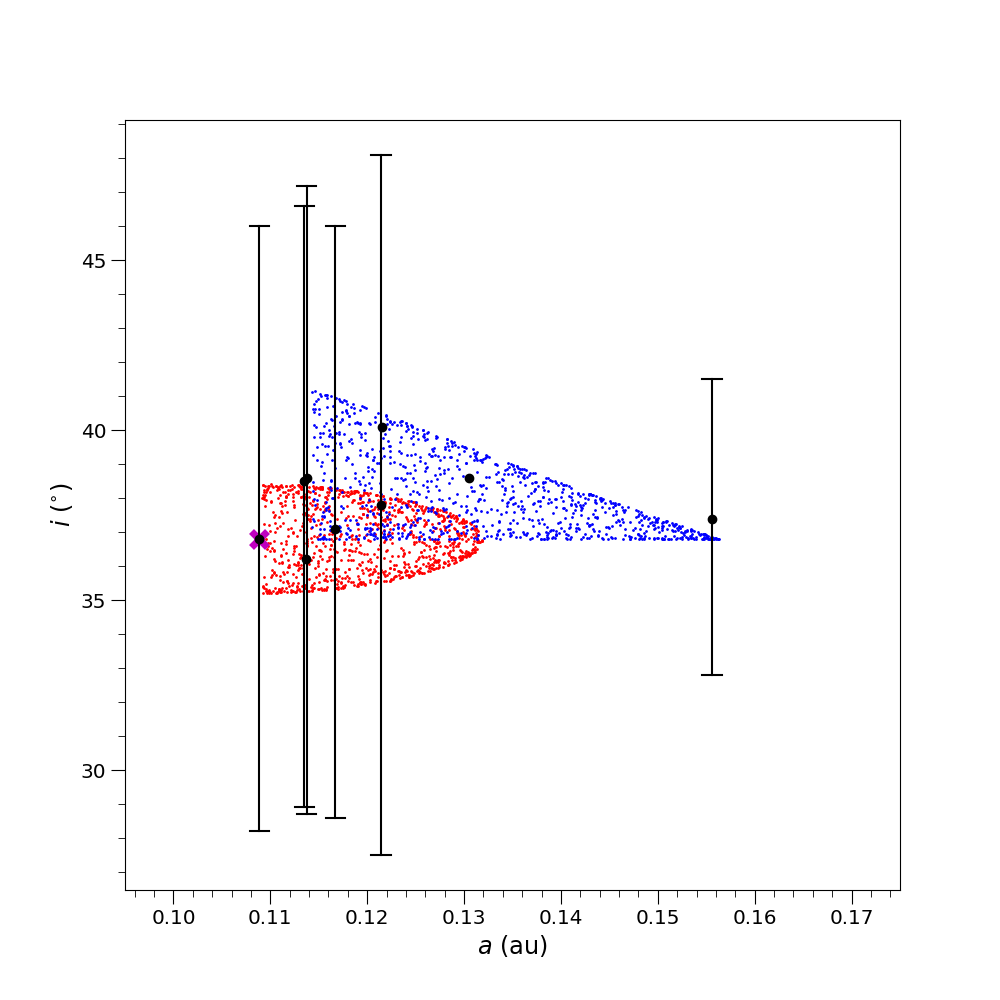}
\caption{A thousand ejecta particles were simulated in attempt to recreate the spread in orbit of the Gallic group with S/2004 S 24 and without. The ejecta particles were created by adding $\Delta v$ = 80~m/s (red dots) or 330~m/s (blue dots) to that of the chosen progenitor (magenta cross). 
Time-averaged orbital elements from JPL of all of the moons in the subgroup are shown by black circles.
The vertical bars through some the real moons indicate the range of orbital values that a moon oscillates over during a secular cycle, taken from \citet{Jacobson2022}. The variations in $a$ are typically comparatively small and are thus not shown. The orbital elements ranges of the newest moons are not shown since they were announced after the \citet{Jacobson2022} study and are therefore not provided by JPL. }
\label{fig:alsim}
\end{figure}

As mentioned in \autoref{sec:gallic}, the Gallic group could be created by a head-on cratering event near the apex point of Albiorix. Thus, for the orbit of the progenitor moon, we used the averaged $a$, $e$, and $i$ of Albiorix, 0.1092~au, 0.48, and 36.8$^{\circ}$ respectively. To crudely simulate a head-on cratering collision we restricted the ejecta to having outbound trajectories that are within 90$^{\circ}$ of the direction of motion of the progenitor. To recreate the comparatively high-$a$ low-$e$ orbit of S/2004 S 24, we set the collision to occur while the progenitor was at apocenter (with true anomaly $f$ = 180$^{\circ}$). The longitude of ascending node, $\Omega$, argument of periapsis, $\omega$, were set to 0$^{\circ}$and 90$^{\circ}$, respectively, to match the small inclination range of the group and the fact that the inclinations of almost all members are greater than that of Albiorix. In order to get ejecta out to the distance of S/2004 S 24 one needs a $\Delta v$ of 330~m/s (see \autoref{fig:alsim}).
This velocity is significantly higher than what is expected for an irregular moon collisional family (see \citet{Turrini2008} and \autoref{sec:veldis}).
Additionally, although the ejecta overlap with the $e$ ranges of all Gallic group members (that have ranges), there appears to be a trend (sans S/2004 S 24) that the larger $a$ members also have slightly larger $e$.

We next turn our attention to just the Albiorix subgroup (this is the Gallic group without S/2004 S 24). To get an ejecta cloud that has increasing $e$ with increasing $a$, we changed $\omega$ to 120$^{\circ}$ and $f$ to 30$^{\circ}$. To reach the outermost member of the Albiorix subgroup, S/2006 S 12, a $\Delta v$ of 80~m/s is needed (see \autoref{fig:alsim}). This much lower $\Delta v$, compared with the Gallic group as a whole, is well within the acceptable range for a collisional family (see \citet{Turrini2008} and \autoref{sec:veldis}). Thus, the Albiorix subgroup itself is a good candidates for a collisional family.

As for the origin of S/2004 S 24, perhaps there was a secondary collision where one of the larger $a$ Albiorix subgroup members was impacted while near apocenter and S/2004 S 24 is the largest fragment of that collision. Alternatively, as suggested by \citet{Sheppard2023}, S/2004 S 24 might not be related to the rest of the Gallic group and, coincidentally, happen to have similar $i$. 
Discovering more Gallic group members should lead to a better understanding of the origin of S/2004 S 24.

\subsection{Mundilfari subgroup}

As mentioned in \autoref{sec:sizedist} there is evidence that the entire Mundilfari subgroup was created by a recent catastrophic collision. To test this, we used the approximate centre of the Mundilfari subgroup for $i$, $a$, and $e$ as the orbit of the progenitor, with values of 165$^{\circ}$, 0.126~au, and 0.28 respectively. 
We varied $\Omega$, $\omega$, and $f$ to find an ejecta field that best matched the orbits of the Mundilfari subgroup. Since this subgroup is more dispersed in orbital element space, we have increased the number of simulated ejecta to 2000.

%------------------------------------ FIGURE 6 -------------------------------
\begin{figure}[ht]
\plottwo{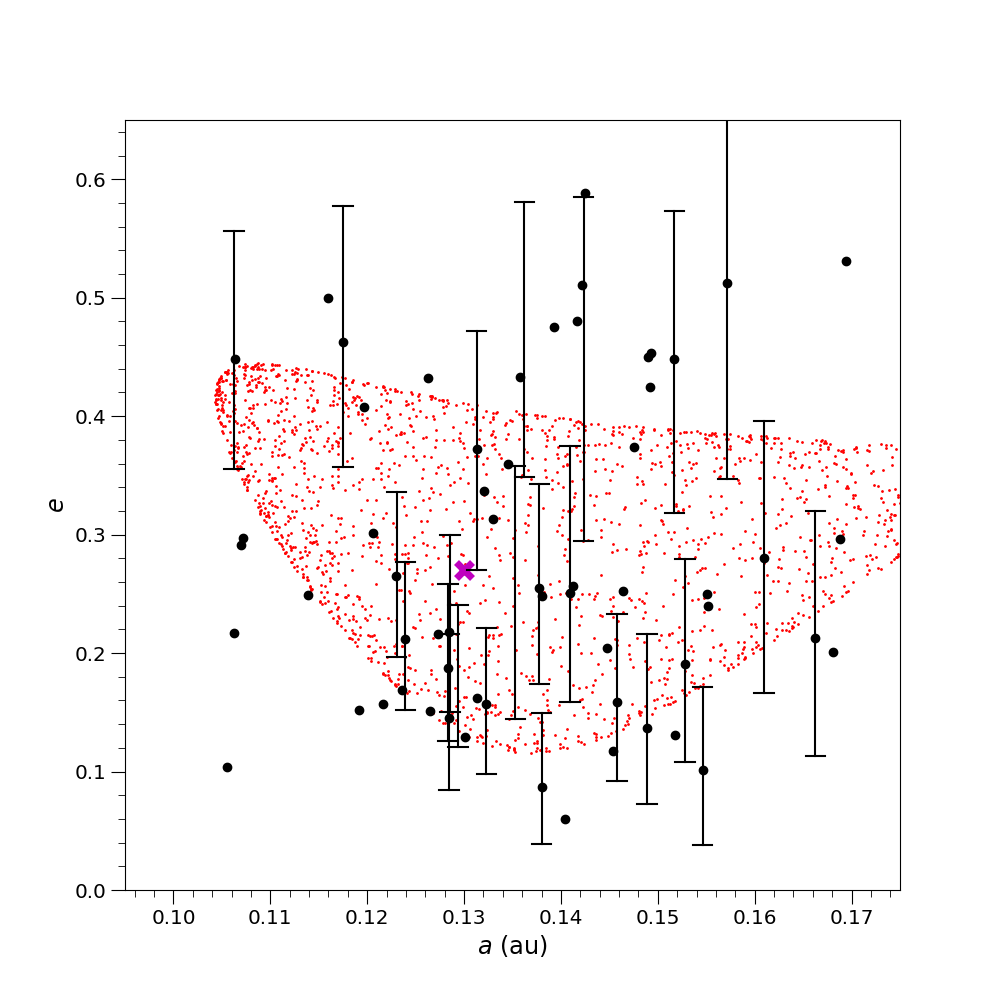}{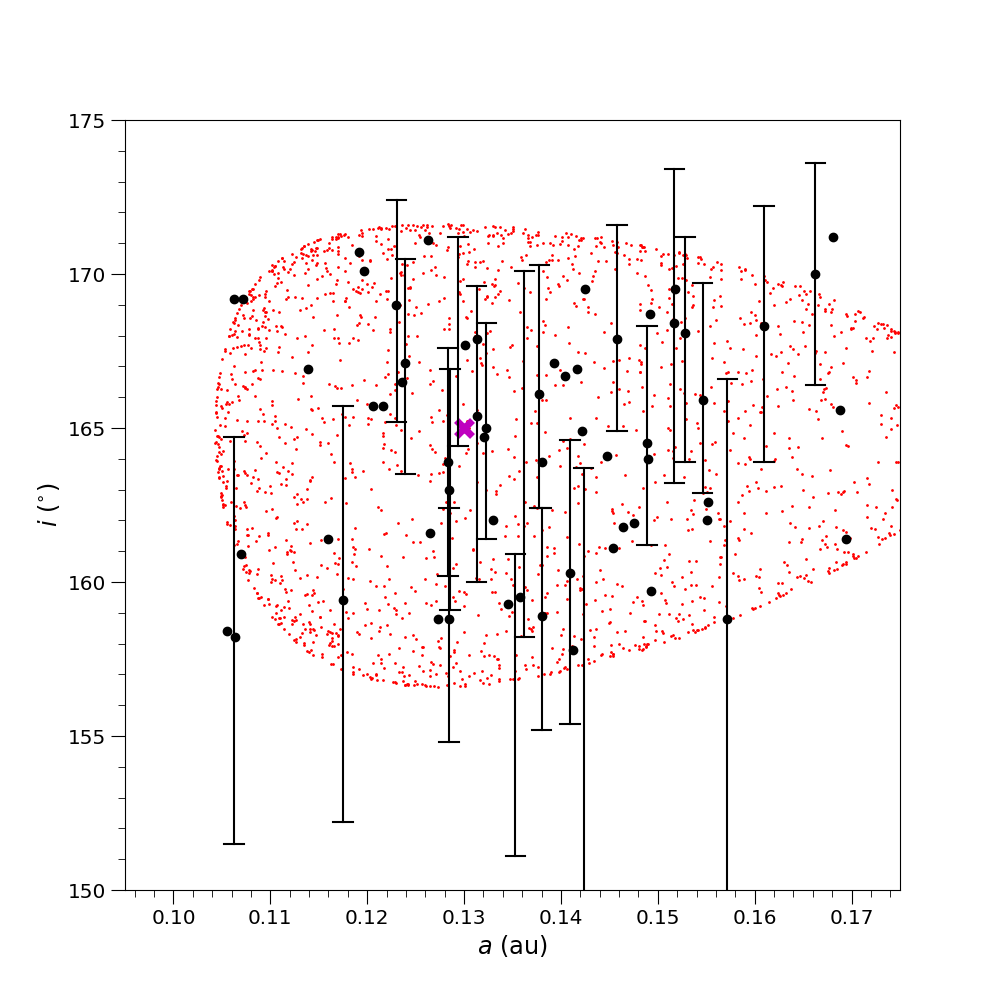}
\caption{An attempt to recreate the spread in orbit of the Mundilfari subgroup members (black) by simulating 2000 ejecta particles (red). The ejecta particles were created by adding $\Delta v$ = 200~m/s in a random direction to that of a chosen progenitor (magenta cross). 
See \autoref{fig:alsim} caption for more details.}
\label{fig:munsin}
\end{figure}
%------------------------------ end Figure 6 ----------------------------------

We find that a minimum of 200~m/s is needed for $\Delta v$ to produce an ejecta field that covers the ranges of almost all of the Mundilfari subgroup members (see \autoref{fig:munsin}). This occurred when the progenitor had $\Omega \simeq 0^{\circ}$, $\omega \simeq 270^{\circ}$, and $f \simeq 240^{\circ}$. It was much easier to fit the spread in $i$ compared to the spread in $e$ and $a$.

\subsubsection{Velocity Dispersion}
\label{sec:veldis}

Based on hydrocode models, \citet{Nesvorny2003} set a velocity dispersion limit of 100~m/s to identify potential collisional families within the saturnian irregular moon population, where the velocity dispersion for a set of moons is obtained using Gauss equations. In \citet{Turrini2008}, they decided to relax the limit to 200~m/s to account for the large oscillations that the orbits of irregular moons can have.
% \citet{Turrini2008} used a velocity dispersion limit of 200~m/s to identify potential collisional families within the saturnian irregular moon population, where the velocity dispersion for a set of moons is obtained using Gauss equations. {\bf Their limit is a relaxation of \citet{Nesvorny2003} limit of 100~m/s to account for the large oscillations that the orbits of irregular moons can have.} [OLD VERSION]
Although our minimum $\Delta v$ is the same as the velocity dispersion limit used by \citet{Turrini2008}, the two velocities are not quite equivalent. Velocity dispersions are calculated using the mean orbital elements, whereas we fit the range of orbital elements. 

The Cluster C family defined in Table 3 of \citet{Turrini2008} contains 11 of 13 Mundilfari subgroup members known then, and additionally Griep, Farbauti, and S/2006 S 1 (all of which have inclinations close to the limits that define the Mundilfari subgroup, see \autoref{sec:sizedist}). Due to the similarities in the members of the Mundilfari subgroup and the Cluster C family, the velocity dispersions of the two sets of moons should be very similar. Since \citet{Turrini2008} found that the Cluster C family has a velocity dispersion of 240~m/s,
the velocity dispersion for the Mundilfari subgroup likely exceeds 200~m/s.
If one accepts this criterion, 
the Mundilfari subgroup appears to be too dispersed to be a collisional family. 

Perhaps some members at the extrema $a$ and/or $e$ of the Mundilfari subgroup were not formed from the recent collision (thus causing the spread in $a$ and/or $e$ of the collisional family to appear larger than it actually is) or perhaps the orbital spread has been enlarged by subsequent collisions within the subgroup but not enough to grind down the steep size distribution. 
Alternately, it could be that there were two (or more) separate collisions that both occurred recently and overlapped slightly in orbital phase space. 
\citet{Li&Christou2018} found that the gravitational interaction with Phoebe causes ejecta from the large moon to disperse over the age of the Solar System. However, the effect of Phoebe on the spread of the Mundilfari subgroup is likely minimal due to the likely recency of the subgroup and, to a lesser extent, the lower rate of Phoebe interactions due to the larger mutual inclinations (than what Phoebe impact ejecta would have).

The Himalia group around Jupiter is another cluster of irregular moons that is a good candidate for a collisional family but again with a velocity dispersion that appears too high \citep{Nesvorny2003}. 
With more than one candidate collisional family in the irregular moon population having more dispersion in orbital phase space than is expected, perhaps some detail in the understanding of irregular moon collisions is missing.

\subsubsection{Colours}

Numerous works have measured the colours of saturnian irregulars \citep{Grav2003,Buratti2005,Grav&Bauer2007,Graykowski2018,Pena2022},
although with the main focus of comparing saturnian irregulars, along with irregular moons of other giant planets, to other populations of small bodies in the Solar System.
There has been very little attempt to use colour information to identify potential collisional families.
This is mainly due to the difficulty of getting high-quality colours of a large enough sample of saturnian irregulars, caused by the faintness of almost all of the moons coupled with their irregular shapes.

Interestingly, \citet{Grav&Bauer2007} found that Mundilfari has a V-I colour of 0.52 $\pm$ 0.07, significantly bluer than almost any other saturnian irregular. The only saturnian irregular that has similar V-I is Phoebe with 0.64 $\pm$ 0.01.
Thus far, V-I colours (or the equivalent g-i in the Sloan filter system) have not been measured for any other Mundilfari subgroup member. 
Aegir and Fornjot were included in the sample from \citet{Graykowski2018}, but they did not obtain observations in the I filter.
\citet{Pena2022} observed Skoll, Fornjot, and Loge but did not provide $g$-band magnitudes for these three moons.
If the Mundilfari subgroup is indeed a collisional family, then we expect that other members will also exhibit blue V-I colours that Mundilfari is observed to have. 

%------------------------------------ FIGURE 7 -------------------------------
%\begin{figure}[ht]
%\plotone{coree.png}
%\caption{Histogram of mean eccentricity of the core sub-group.}
%\label{fig:histecore}
%\end{figure}
%------------------------------ end Figure 7 ----------------------------------

\section{Conclusion}
\label{sec:con}

Irregular moons have likely undergone significant collisional evolution since being captured early in the Solar Systems history. This evolution can be inferred from the current orbital distribution of an irregular moon population.
From the discovery and precise tracking of 64 new irregular moons of Saturn, increasing the known number twice fold, we are able to accurately divide the known groups into subgroups. 
In the Inuit group, we identify two tightly packed subgroups, the Kiviuq and Siarnaq subgroups, each of which are likely to be collisional families.
Using the inclination distribution, we split the Norse group into 3 subgroups, which we refer to as the Phoebe, Mundilfari, and Kari subgroups.

The Mundilfari subgroup, with differential size index $q$=6, has a significantly steeper size distribution slope compared to all other groups/subgroups. 
Due to the steep size distribution, we believe that the Mundilfari subgroup was created by the recent collision proposed by \citet{Ashton2021}.
A major concern, however, is the large orbital phase space that the Mundilfari subgroup occupies, which is larger than one might expect for the spread coming from the break-up speed of a collisional family.
The recent collision hypothesis would be greatly strengthened if other members of the Mundilfari subgroup were found to have the very blue V-I colour of Mundilfari.

\section{Acknowledgments}

This work was supported by funding from the Natural Sciences and Engineering Research Council of Canada. Thanks to the CFHT Queue Observing team, especially Todd Burdullis and Daniel Devost, for helping us with the data acquisition process. This research used the facilities of the Canadian Astronomy Data Centre operated by the National Research Council of Canada with the support of the Canadian Space Agency.

%\appendix

%All of our measurements have been submitted to the Minor Planet Center.

\clearpage

\bibliographystyle{aasjournal}
\bibliography{newmoons}{}

%% This command is needed to show the entire author+affilation list when
%% the collaboration and author truncation commands are used.  It has to
%% go at the end of the manuscript.
%\allauthors

%% Include this line if you are using the \added, \replaced, \deleted
%% commands to see a summary list of all changes at the end of the article.
%\listofchanges

\end{document}